# The LVD Core Facility: a study of LVD as muon veto and active shielding for dark matter experiments

**Marco Selvi, on behalf of the LVD collaboration**

*INFN - Bologna*
*Via Irnerio 46, Bologna - Italy*
*E-mail:* `selvi@bo.infn.it`

Many experiments looking for dark matter are aiming to get the ton scale in the future. However, it is well known that scaling dark matter detectors to higher mass is not a sufficient condition for sensitivity and that an equally important condition is to simultaneously keep the background low, in particular the ultimate background, namely the fast neutron background induced by muons. In this study we explore the possibility of using the existing structure of a running experiment, the LVD supernova observatory at the INFN Gran Sasso National Laboratory, as an active shield and veto for the muon-induced background.

In our vision LVD could become (without affecting in any way its main purpose of SN neutrino telescope) a 'host' for a relatively compact but massive experiment looking for rare events. The LVD experiment consists of a 1000 ton liquid scintillator detector with a highly modular structure, being made of 3 identical towers, each one composed by 35 active modules. The empty volume that can be obtained removing 2 modules from the most internal part of the detector is 2.1m x 6.2m x 2.8m; we will call it "LVD Core Facility" (LVD-CF).

We have evaluated the active vetoing and shielding power of LVD, with a detailed MC simulation (based on Geant4) of the detector and the rock that surrounds it. We have generated cosmic muons with energy spectrum and angular distribution sampled accordingly to what is expected in the LNGS underground laboratory ($<E>$=270 GeV). The number and energy spectrum of the muon-induced neutrons that enter the LVD-CF has been calculated. The results show that the flux of neutrons that are not associated with a visible muon in LVD is very low; it results reduced by a factor 50, equivalent to the one present in a much deeper underground laboratory, i.e. Sudbury. Moreover we present the results of on-going measurements about the gamma contamination inside the LVD-CF: it is reduced by a factor greater than 10 with respect to the one measured outside the LVD detector.

## 1. Introduction

The overall framework of Dark Matter searches is of an increasing competition. In particular the Gran Sasso laboratory is the one with the highest concentration of Dark Matter experiments, in quantity and quality. Within 2008 two of the existing experiments will scale up their sensitive mass to the 100 kg scale, having already started design studies for the next generation experiments at the ton scale. This will maintain the leading role of the Gran Sasso laboratory as the site for the most sensitive dark matter experiments worldwide. At the same time it is clear that even the Gran Sasso laboratory cannot provide adequate space for all next generation experiments in the field of Dark Matter search. Furthermore, the sensitivity goal of future Dark Matter searches is such that greater depths than that of Gran Sasso are required.

It is well known that scaling dark matter detectors to higher mass is not a sufficient condition for sensitivity and that an equally important condition is to simultaneously keep the background low, in particular the ultimate background, namely the fast neutron background induced by muons. Next generation experiments will require deep sites and active shields with masses of the order of 1 kt.

The proposal to use the existing structure of a running experiment, the LVD supernova observatory at Gran Sasso, as an active shield and muon veto for a next generation dark matter experiment addresses specifically these issues and appears to provide an ideal solution. In our vision, LVD could become (without affecting in any way its main purpose of SN neutrino telescope) a 'host' for a relatively compact but massive experiment looking for rare events [1].

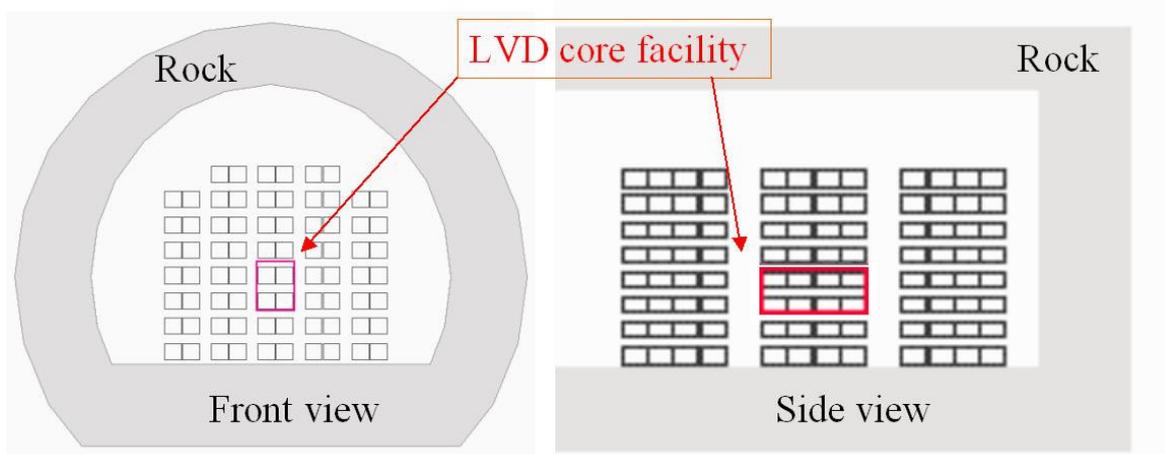

*Figure 1: front and side view of the LVD detector and its position relative to the rock around it. The red box represents the LVD Core Facility.*

## 2. The proposed concept

The LVD detector [2,3] (1000 ton of liquid scintillator) has a highly modular structure, being made of 3 identical towers, each one composed by 35 active modules. Each module ("portatank") contains 8 scintillation counters, 1.5 m$^3$ each. The removal of two modules of the innermost region of the experiment, makes available a volume of 2.1m x 6.2m x 2.8m (height), see figure 1, which we name "LVD Core Facility" (hereafter LVD-CF or simply CF). As we will show, this volume, which is the minimum required for an 1-ton dark matter experiment, is probably the most background-free of the entire Gran Sasso laboratory.

## 3. The muon-induced neutron flux

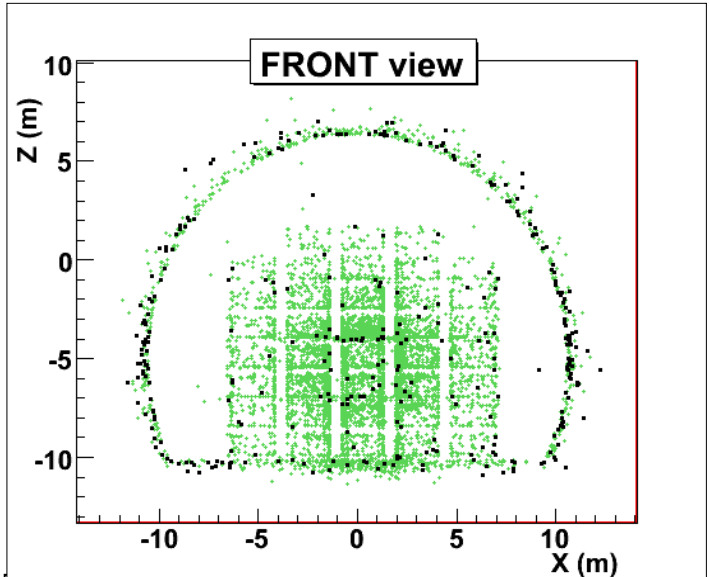

*Figure 2: Front view of the positions where the neutrons that enter the LVD-CF are produced. In green those neutrons whose parent muon is tagged, while in black the untagged*

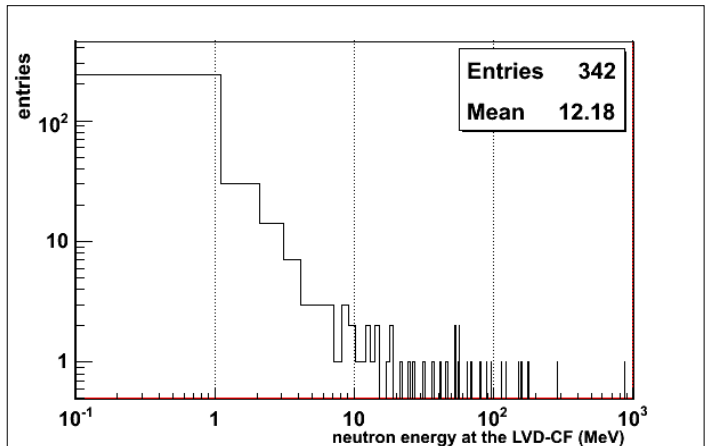

*Figure 3: The energy spectrum of the 342 neutrons which enter the LVD-CF, not associated with a tagged muon.*

It is well known that the muon-induced fast neutron background is the ultimate limitation to the sensitivity of rare event searches, such as dark matter searches.

While it is relatively easy to stop neutrons up to about 10 MeV with shields of reasonable thickness, the more energetic ones need a different approach with active veto systems.

We have evaluated the active vetoing and shielding power of LVD, with a detailed MC simulation (based on Geant4) of the detector and the rock that surrounds it. We have generated cosmic muons with energy spectrum and angular distribution sampled accordingly to what is expected in the LNGS underground laboratory (<E>=270 GeV). The number and energy spectrum of the muon-induced neutrons that enter the LVD-CF has been calculated.

Because of its large volume, LVD can detect muons even very far from the CF: the LVD external dimensions, indeed, are 13m x 22.7m x 10m (LVDbox). Nevertheless inside LVD there are gaps and corridors, thus some muons can cross this volume without being detected.

A "tagged" muon is defined by the time coincidence of at least two scintillation counters. About 85% of the muons that hit the LVDbox are tagged.

We generated 5 10$^6$ muons (corresponding to about 8 months of data acquisition): the number of neutrons that enter the LVD-CF when the parent muon is untagged is just 342. Only 25% of them come from muons that cross the LVDbox and go through a corridor; the remaining 75% are neutrons produced in the rock around the detector that go inside the CF mainly through one of the gaps, as can be seen in figure 2.

The energy spectrum of the neutrons from untagged muons is shown in Fig. 3. The number of neutrons with energy larger than 10 MeV is 36, i.e. about one every 7 days. In terms of flux this corresponds to 2.3 10$^{-12}$ cm$^{-2}$ s$^{-1}$.

The maximum delay between the neutron arrival time and the parent muon, for $E_n$ > 10 MeV, is 400 ns. Thus the dead time introduced when using LVD as a veto is well below 1%.

The importance of LVD as veto and shield emerges clearly if we repeat the MC simulation considering the CF without the LVD detector around it. For $E_n$ > 10 MeV, the number of neutrons in the CF, normalized to the same statistics (8 months) is about 50 times higher than the number of neutrons from untagged muons when the LVD detector is considered.

A similar MC simulation has been developed for the Sudbury site, considering in details its depth, rock composition and cosmic muon flux. The resulting muon-induced neutron flux is comparable to what we get in the LVD-CF, using LVD as a veto. This result makes the muon-induced neutron background in the LVD Core Facility equivalent to that of the deepest existing underground laboratories, i. e. Sudbury.

A summary of the results from our MC simulations are in the following table:

| n flux: 10$^{-9}$ cm$^{-2}$ s$^{-1}$ | LNGS Hall | LVD passive | LVD as µ veto | Sudbury Hall |
|---|---|---|---|---|
| **Total** | 1.78 | 0.60 | 0.0220 | 0.0337 |
| **En > 1 MeV** | 0.30 | 0.30 | 0.0066 | 0.0048 |
| **En > 10 MeV** | 0.11 | 0.10 | 0.0023 | 0.0015 |
| **En > 100 MeV** | 0.03 | 0.03 | 0.0005 | 0.0004 |

**4. Gamma ray and neutrons from spontaneous fission and (α,n) reactions**

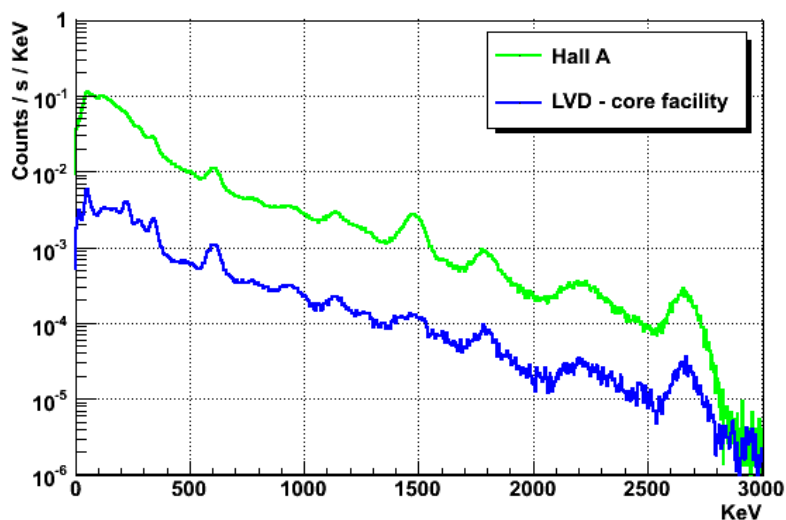

*Figure 4: Gamma ray energy spectrum measured inside the LVD-CF (blue line) and in the hall A outside LVD (green line)*

Another fundamental issue to assess the appropriateness of the LVD-CF as a low-activity site is, of course, the gamma-ray background. A first feeling of this aspect has been obtained by a set of preliminary measurements carried out with a portable 2-inch NaI detector.

As shown in Figure 4, the gamma ray intensity inside the LVD-CF is reduced by a factor greater than 10 with respect to the one measured outside the LVD detector.

This first, encouraging result is telling us that the overall structure of the experiment (which is made out of 1kt of iron) is by itself rather clean, and acts as a shield against the gamma-ray flux coming from the Hall A walls.

Nevertheless, a dark matter experiment, or any experiment searching for rare events, will most likely need a further shielding against gammas and low energy neutrons.
In order to assess the characteristics of this shield, some further studies are needed, namely:

- a more detailed gamma-ray mapping of the LVD-CF (by measurements);

- a more detailed simulation of the surroundings of the LVD-CF that will use data coming from High Purity Ge measurements of the LVD materials (iron, stainless steel, PMTs and liquid scintillator).

These studies are currently underway. The results will clarify the overall background impacting on the LVD-CF and will dictate the best options for additional shielding, if necessary.

**5.Conclusions**

The idea of creating a low background space inside the LVD experiment at Gran Sasso, by removing two of the innermost 8-tanks module seems to be a concrete possibility, with several advantages from the point of view of physics, of space utilization, of costs. In particular, from the point of view of the fast neutron flux, the LVD-CF is comparable with the deepest experimental sites in the world.
Therefore, we will continue our efforts in this direction, in order to further clarify the relevant issues, and namely:

- a precise assessment of the gamma and spontaneous fission and ($\alpha$, n) background;

- a precise evaluation of the fast neutron flux, including the part generated by the guest detector itself;

- a study of the additional shielding required, if any;

- a study of the logistics and mechanical issues involved in realizing the CF;

- a study of the costs.